\documentclass[dvipsnames,twocolumn]{aastex63}
\PassOptionsToPackage{usenames,dvipsnames}{xcolor}
\usepackage{tabularx}
\usepackage{amsmath}  
\usepackage{amssymb}  
\usepackage[T1]{fontenc}
\usepackage{xspace}
\newcommand*{\byosed}{\textsc{byosed}\xspace}

\submitjournal{ApJ}
\shorttitle{Simulating Spectral Variations with \byosed}
\shortauthors{Pierel et al.}

\newif{\ifchangetext}
\changetextfalse

\InputIfFileExists{\jobname-options}

\ifchangetext
  
  \newcommand{\changenote}[1]{\textcolor{blue}{ \bf #1}}x
\else
  
  \newcommand{\changenote}[1]{}
\fi

\hypersetup{
    colorlinks=True,
    citecolor=black,
    linkcolor=black,
    filecolor=magenta,      
    urlcolor=cyan,
}

\def\xone{\ensuremath{x_{1}}}
\def\C{\ensuremath{\mathcal{C}}}

\newcommand*{\kaepora}{\textsc{kaepora}\xspace}

\defcitealias{hounsell_simulations_2018}{H18}
\newcommand{\hounsell}{\citetalias{hounsell_simulations_2018}}

\defcitealias{oguri_gravitationally_2010}{OM10}

\defcitealias{pierel_turning_2019}{PR19}

\begin{document}

\title{\Large Understanding Type Ia Supernova Distance Biases by Simulating Spectral Variations}

\correspondingauthor{J.~D.~R.~Pierel}
\email{jr23@email.sc.edu}
\author{J.~D.~R.~Pierel}

\affil{Department of Physics \& Astronomy, University of South Carolina, 712 Main St., Columbia, SC 29208, USA}
\author{D.~O.~Jones}

\affil{Department of Astronomy \& Astrophysics, University of California, 1156 High St., Santa Cruz, CA 95064, USA}
\affil{NASA Einstein Fellow}
\author{M.~Dai}

\affil{Department of Physics \& Astronomy, Johns Hopkins University, 3701 San Martin Drive, Baltimore, MD 21218, USA}
\affil{Department of Physics \& Astronomy, Rutgers, the State University of New Jersey, Piscataway, NJ 08854, USA}
\author{D.~Q.~Adams}

\affil{Department of Physics \& Astronomy, University of South Carolina, 712 Main St., Columbia, SC 29208, USA}
\author{R.~Kessler}

\affil{The Kavli Institute for Cosmological Physics, Chicago, IL 60637, USA}
\author{S.~Rodney}

\affil{Department of Physics \& Astronomy, University of South Carolina, 712 Main St., Columbia, SC 29208, USA}
\author{M.~R.~Siebert}

\affil{Department of Astronomy \& Astrophysics, University of California, 1156 High St., Santa Cruz, CA 95064, USA}
\author{R.~J.~Foley}

\affil{Department of Astronomy \& Astrophysics, University of California, 1156 High St., Santa Cruz, CA 95064, USA}
\author{W.~D.~Kenworthy}

\affil{Department of Physics \& Astronomy, Johns Hopkins University, 3701 San Martin Drive, Baltimore, MD 21218, USA}
\author{D.~Scolnic}

\affil{Department of Physics, Duke University Durham, NC 27708, USA}

\label{sec:abstract}
\begin{abstract}
In the next decade, transient searches from the Vera C. Rubin Observatory and the \textit{Nancy Grace Roman Space Telescope} will increase the sample of known Type Ia Supernovae (SN Ia) from $\sim${}$10^3$ to $10^5$. With this reduction of statistical uncertainties on cosmological measurements, new methods are needed to reduce systematic uncertainties. Characterizing the underlying spectroscopic evolution of SN Ia remains a major systematic uncertainty in current cosmological analyses, motivating a new simulation tool for the next era of SN Ia cosmology: Build Your Own Spectral Energy Distribution (\byosed). \byosed is used within the SNANA framework to simulate light curves by applying spectral variations to model SEDs, enabling flexible testing of possible systematic shifts in SN Ia distance measurements. We test the framework by comparing a nominal \textit{Roman} SN Ia survey simulation using a baseline SED model to simulations using SEDs perturbed with \byosed, and investigate the impact of neglecting specific SED features in the analysis. These features include semi-empirical models of two possible, predicted relationships: between SN ejecta velocity and light curve observables, and a redshift-dependent relationship between SN Hubble residuals and host galaxy mass. We analyze each \byosed simulation using the SALT2 \& BBC framework, and estimate changes in the measured value of the dark energy equation-of-state parameter, $w$. We find a difference of $\Delta w=-0.023$ for SN velocity and $\Delta w=0.021$ for redshift-evolving host mass when compared to simulations without these features. By using \byosed for SN Ia cosmology simulations, future analyses (e.g., Rubin and Roman SN Ia samples) will have greater flexibility to constrain or reduce such SN Ia modeling uncertainties.
\end{abstract}

\section{Introduction}
\label{sec:intro}
Type Ia Supernovae (SN Ia) are used for one of the fundamental observational tests of the $\Lambda$CDM ``concordance cosmology.''  
The accelerating expansion of the universe was discovered with a sample of a few dozen SN Ia extending to $z\sim0.8$ \citep{riess_observational_1998,perlmutter_measurements_1999}, and over the last 20 years this sample has grown to include $>$1000 SN Ia reaching to $z\sim2.3$ \citep{scolnic_complete_2018}. SN Ia are not standard candles, in that they do not all
have the same absolute brightness. Rather, they are 
standardizable candles.  Cosmological constraints depend upon SN Ia luminosity distance measurements, obtained by fitting the observed light curves with a carefully trained model \citep[e.g.,][]{guy_salt2:_2007,jha_improved_2007}. The model is used to standardize SN Ia absolute magnitudes by correcting for luminosity dependencies on model parameters such as shape and color \citep[e.g.,][]{tripp_two-parameter_1998}. To test the SN Ia training and standardization process, 
and to evaluate robust systematic uncertainties, here we introduce \textit{Build Your Own Spectral Energy Distribution} (\byosed)\footnote{\href{https://github.com/jpierel14/BYOSED}{https://github.com/jpierel14/BYOSED}  \href{https://byosed.readthedocs.io/en/latest}{https://byosed.readthedocs.io/en/latest}}, a Python utility 
to generate realistic SN Ia SED models for simulated 
data samples.  

Due to uncertainty in the physics of SN Ia progenitors, standardization of SN Ia is entirely empirical. For the last decade, the most commonly used and scrutinized model that implements shape and color corrections is the SALT2 model \citep{guy_salt2:_2007,guy_supernova_2010}. SALT2 has been instrumental in SN Ia cosmology analyses due to its accuracy in reproducing observed SN Ia light curves, and has become the standard choice for both large-scale SN Ia simulations and light curve fitting for luminosity distance measurements \citep[e.g.,][]{betoule_improved_2014,scolnic_measuring_2016,jones_should_2018,scolnic_complete_2018,abbott_first_2019,jones_foundation_2019,kessler_first_2019,smith_first_2020}. 

As more data have become available, additional observables have been identified that correlate with SN Ia properties or distance measurements. One of the earliest observations of such a relationship was based on the total stellar mass of the SN host galaxy \citep[e.g.,][]{kelly_hubble_2010,sullivan_dependence_2010,lampeitl_effect_2010}. More recent studies suggest that corrections based on galaxy properties at the SN location could lead to more accurate SN Ia distance measurements \citep{rigault_evidence_2013,kim_type_2014,jones_should_2018,rigault_strong_2018,roman_dependence_2018,jha_observational_2019}. Today, it is generally accepted that the environment of a SN Ia correlates with the properties of its spectra and light curve, but it is unclear if current corrections (e.g., the mass step) account for these relationships with sufficient precision for much larger data sets that will be obtained over the next decade. Numerous studies have investigated the potential impact on SN Ia standardization of many variables beyond host mass including local specific star formation rate \citep[sSFR,][]{rigault_evidence_2013,kelly_distances_2015}, intrinsic luminosity evolution with redshift \citep{howell_predicted_2007}, and metallicity \citep{childress_host_2013,hayden_fundamental_2013}, as well as the ways in which SN properties such as ejecta velocity correlate with Type Ia standardization parameters and Hubble residuals \citep{foley_velocity_2011,siebert_possible_2020}.

At present, simulations of SN Ia populations are used in cosmological analyses to correct for predicted or empirically measured biases \citep[e.g.,][]{betoule_improved_2014,scolnic_measuring_2016,kessler_correcting_2017,scolnic_complete_2018,kessler_first_2019}, and to plan for future surveys \citep[e.g.,][]{hounsell_simulations_2018,kessler_models_2019}. In all cases, the simulations are created by drawing random stretch and color parameters from measured population distributions to obtain variations of the SALT2 model. Current simulations for bias corrections are capable of including relationships between SN Ia properties and the SN host environment at the population level, but are not flexible enough to apply SED features and their correlations with stretch, color,
and host galaxy properties directly. This modeling limitation results in inadequate tests for biases in measurements of SN Ia distances and cosmological parameters.

\byosed provides a new and general method for producing simulated SN Ia light curve samples, by using real SN Ia spectra or a theoretical model to simulate these potential effects on a SN Ia population. \byosed is uniquely poised to help identify inadequacies in SN Ia light curve models such as SALT2, by independently simulating variations beyond the scope of the model used to fit the generated light curves. Previous simulated bias corrections for the JLA \citep{betoule_improved_2014}, Pantheon \citep{scolnic_complete_2018}, PS1 \citep{jones_measuring_2018}, and DES3YR \citep{abbott_first_2019} have used the SALT2 model and the SuperNova ANAlysis \citep[SNANA:][]{kessler_snana:_2009,kessler_results_2010,kessler_first_2019} software package. These analyses do not characterize the SED features correlating with observables such as host galaxy mass or SN ejecta velocity, which could propagate through to distance measurements and cosmological inferences. \byosed also generates SN Ia light curves within the SNANA framework for state-of-the-art bias-correction and cosmological parameter estimation, but is able to include these features directly into the source SED. 

There has been one previous effort to characterize SALT2 training uncertainties with a more general simulation \citep{mosher_cosmological_2014}. This effort focused on warping SED templates from \citet{hsiao_k_2007} in order to avoid generating a model in which the SALT2 spline basis is already imprinted on the SEDs. However, \citet{mosher_cosmological_2014} did not introduce correlations between the host galaxy and the SN SED, nor did they introduce redshift-dependent parameters.

Section \ref{sec:framework} gives a description of the implementation of \byosed, including the process needed to vary a baseline SED with observed spectra. Section \ref{sec:sims} provides a simulation case study in which light curves are generated with \byosed inside of SNANA, and fit to obtain distance measurements. These measurements are propagated to Section \ref{sec:biases} to identify resulting biases in the dark energy equation of state parameter, $w$, relative to a fiducial survey. We conclude with a discussion of the results in section \ref{sec:discussion}.

\begin{figure*}[t!]
    \centering
    \includegraphics[trim={1cm 0 1cm 0cm },clip,width=\textwidth]{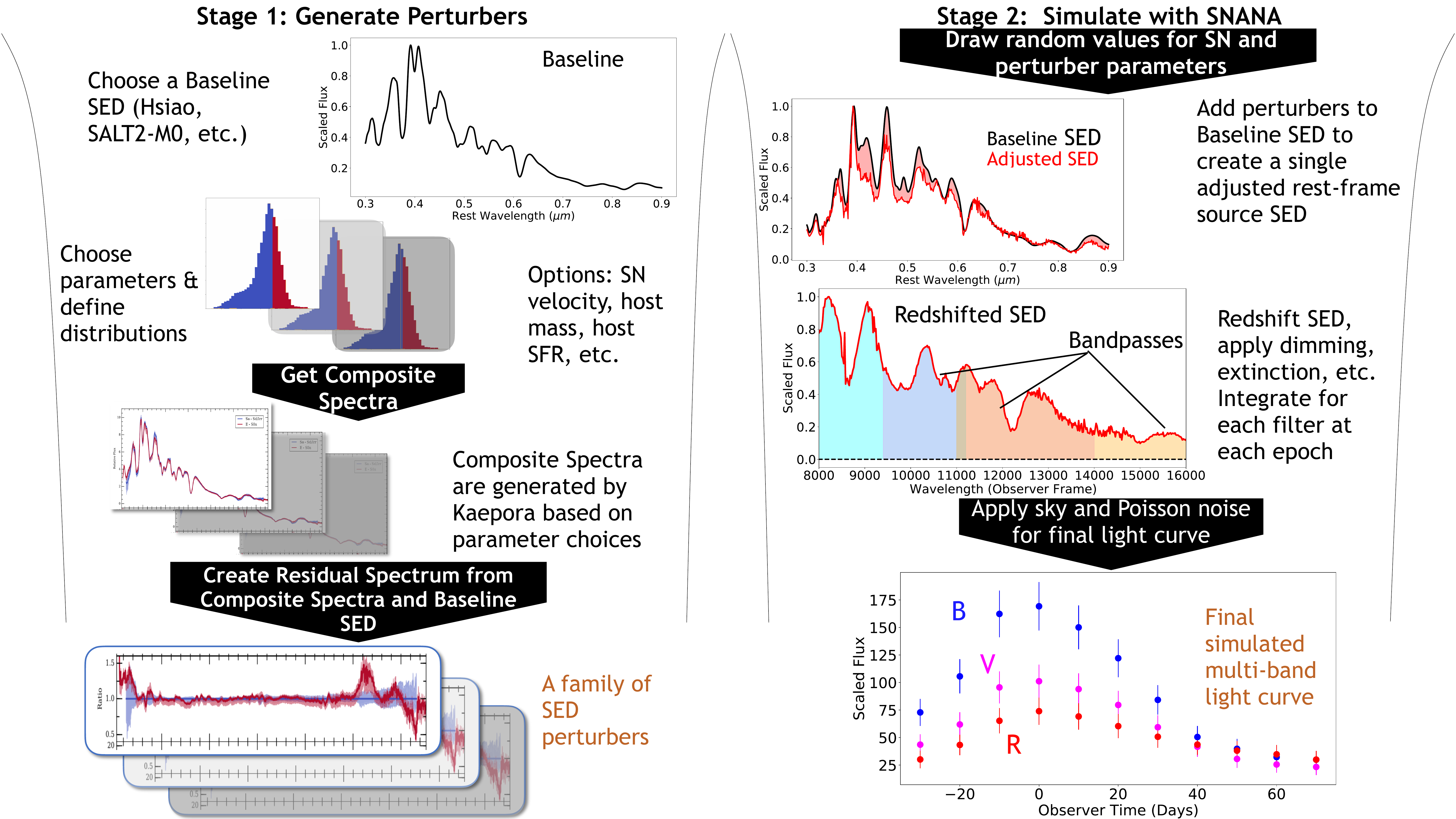}
    \caption{A simulation using \byosed proceeds in two stages. First (left) perturbers are created using \kaepora spectra, along with the associated parameter distributions used to create (for example) high and low ejecta velocity SN Ia. Next (right), the perturber parameter(s) are set (e.g., ejecta velocity) and the corresponding perturber is used to create a redshifted SED for light curve simulations inside of SNANA.}
    \label{fig:byosed_schematic}
\end{figure*}

\

\section{BYOSED}
\label{sec:framework}
\byosed is a software tool designed to produce flexible representations of SN Ia light curves by varying a baseline SED model used by the SNANA simulation. The process of simulating SN light curves with SNANA is summarized by Figure 1 in \citet{kessler_first_2019}, which begins with a rest-frame ``Source SED'' and then propagates the SN light through an expanding universe, the Milky Way Galaxy, Earth's atmosphere, instrumental filters, and finally generates CCD photoelectrons. \byosed fills the Source SED step in an SNANA simulation, preserving all
of the detailed instrumental and cadence modeling that follows to produce realistic SN Ia light curves for any survey. Figure \ref{fig:byosed_schematic} provides a graphical representation of the process described below from baseline SED to light curve simulations with \byosed and 
SNANA. Fits to these light curve simulations do not account for differences at the spectral level between the \byosed source SED and the light curve fitting model, which could manifest themselves as biases on distance measurements for each SN Ia. By propagating these fitted distances through to cosmological measurements, we can estimate any cosmology parameter biases that may arise. In this work, we present example case studies of this entire process from building a \byosed source SED to identifying potential biases for cosmology.

\begin{figure*}[t!]
    \centering
    \includegraphics[width=\textwidth]{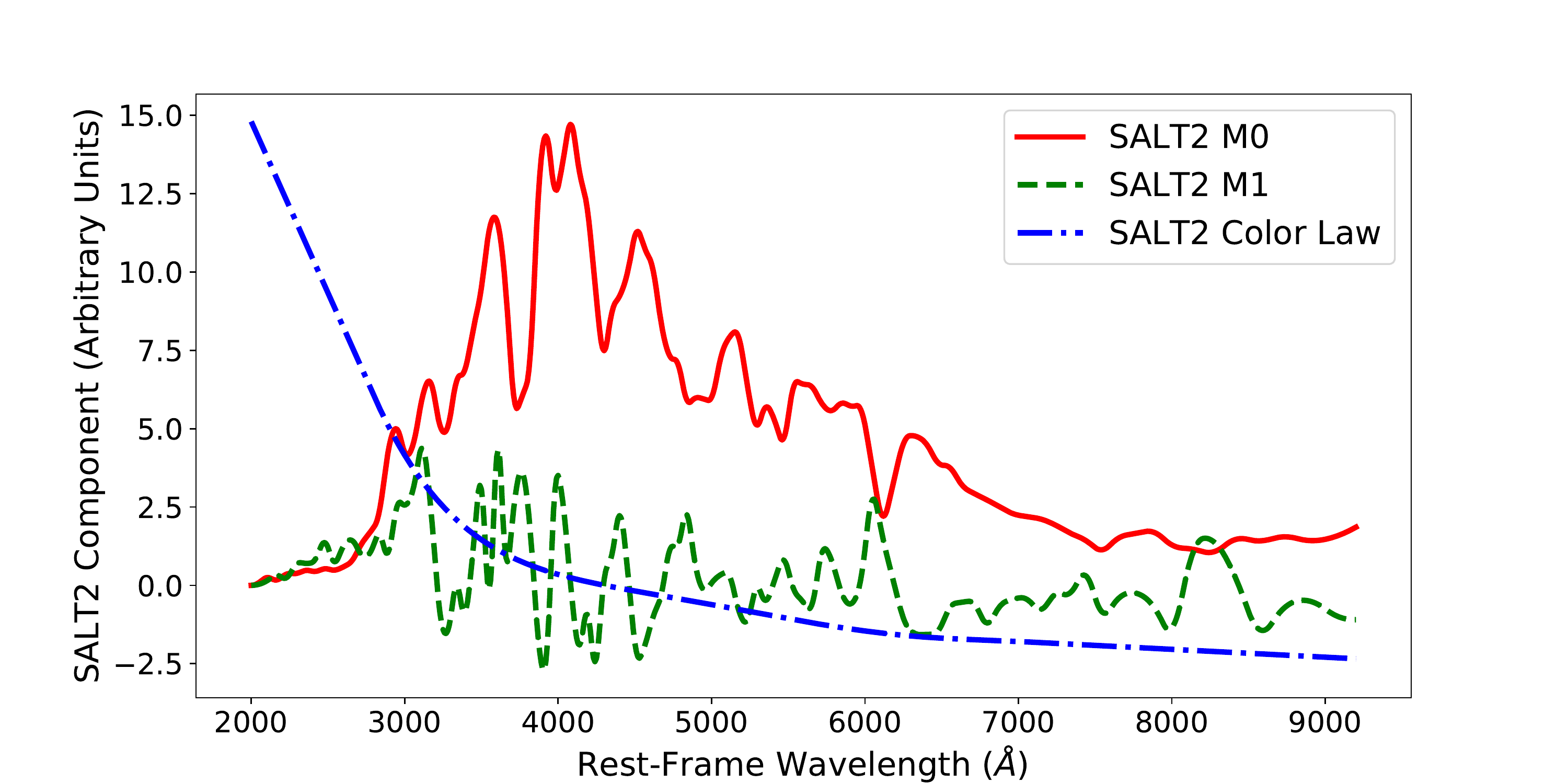}
    \caption{The primary components that make up the existing SALT2 model at peak B-band brightness. The $M_0$ component (red solid) defines a ``baseline'' SN Ia spectrum, the $M_1$ component (green dashed) controls the stretch of a given SN, and the color-law (blue dash-dot) is equivalently $A_\lambda-A_{\rm{B}}$.}
    \label{fig:salt2}
\end{figure*}

\subsection{Framework}
\label{sub:salt2}
\byosed is independent of any single SN Ia SED model, but it is instructive to begin with a description of the commonly used {\it Spectral Adaptive Light-curve Template 2} (SALT2; \citealt{guy_salt:_2005,guy_salt2:_2007,guy_supernova_2010}) for comparison and context. SALT2 is a parametric light curve model first developed for analysis of the Supernova Legacy Survey data (SNLS; \citealt{astier_supernova_2006,guy_supernova_2010,conley_supernova_2011}), and is most accurately trained in the optical wavelength range \citep{betoule_improved_2014,scolnic_systematic_2014,pierel_extending_2018}. SALT2 gives the SN~Ia flux density $F$ as a function of time relative to peak brightness (phase, $t$) and wavelength $\lambda$:
\begin{equation}
\label{eq:salt2}
F(t,\lambda)=x_0\big[M_0(t,\lambda)+x_1M_1(t,\lambda)\big]\exp\big[c\times CL(\lambda)\big] .
\end{equation}
The core components of the SALT2 model ($M_0$, $M_1$, and the color law [$CL$]) are determined in the training process, and are fixed for
every SN Ia (Figure \ref{fig:salt2}). For each SN Ia, a light curve fit is used to determine the
amplitude ($x_0$), stretch parameter ($x_1$), color parameter ($c$),
and time of peak brightness ($t_0$). These parameters are used to standardize the SN Ia brightness and measure a luminosity distance using the Tripp equation \citep{tripp_two-parameter_1998}.

The past decade has shown increasing evidence for SN Ia brightness dependence beyond the simple stretch and color relationship used in the SALT2 framework. Some new models that aim to more accurately capture SN Ia spectral evolution have been proposed \citep[e.g.,][]{saunders_snemo_2018,leget_sugar_2020}, but it is still unclear the extent to which these address current gaps in the SALT2 model. At present, SN Ia distance biases are corrected with simulations that use the SALT2 model to generate the ``observed'' light curves, and subsequently fit each light curve with the same SALT2 model. While Malmquist biases are well modeled, this simulation strategy cannot predict biases from inadequate modeling of the SN Ia SED, and therefore current cosmology constraints are missing this important systematic effect. 

\byosed is designed to precisely address these limitations in the cosmology analysis pipeline. Simulations using \byosed can be used to both identify and quantify presently unaccounted for biases, and determine the relative benefit of using one model over another for cosmological analyses. The two core components of the \byosed framework are a ``baseline SED'' and one or more ``perturbers'', defined as follows:

\begin{enumerate}
    \item \textbf{Baseline SED:} An SED model describing the evolution of a SN Ia as a function of wavelength and time, $H(t,\lambda)$. Examples include the SALT2 $M_0$ component or the template from \citet{hsiao_k_2007} (Hereafter H07).
    \item \textbf{Perturber:} A relationship ($P_i$) derived from theory or observations, which \byosed uses to ``perturb'' the baseline SED model with added SED features that correlate with observables. Examples include spectra defining SN Ia with high and low ejecta velocity, or a correlation between host galaxy metallicity and SN Ia luminosity.
\end{enumerate}

The process used to create these perturbers is described below in Section \ref{sub:perturbers}. With a set of perturbers, \byosed produces a rest-frame source SED:
\begin{equation}
\label{eq:byosed}
\begin{aligned}
	F(t,\lambda)=A\times H(t,\lambda)\Big[1+P_1(\theta_{SN},\theta_{HOST},t,\lambda)s_1+...\\
	+P_N(\theta_{SN},\theta_{HOST},t,\lambda)s_N\Big]\times\\ C_1(\theta_{c_1})c_1\times...\times C_M(\theta_{c_M})c_M,
\end{aligned}
\end{equation}

\noindent where $\theta_{SN}$ is a set of SN parameters (e.g. ejecta velocity), $\theta_{HOST}$ is a set of host galaxy parameters (e.g. mass, metallicity, redshift, etc.), and $s_i$ is a scale factor for $P_i$, the $i^{\rm{th}}$ SED perturber. Here $C_j$ and $c_j$ are the $j^{th}$ ``global correction'' and scale parameter, respectively. A global correction is simply a perturber that corrects the overall flux of the model, similar to the color law in Equation \ref{eq:salt2}, which is dependent on parameters $\theta_{C_i}$. Note that it is possible to create a SALT2-like simulation with \byosed by choosing stretch and color perturbers (see Section \ref{sec:sims}) that mimic the observed variation in SN Ia light curve shape and color described by the SALT2 $M_1$ and CL components (Equation \ref{eq:salt2}). In this case, the \byosed framework would create a rest-frame SED, $F(t,\lambda)$, in the following way:

\begin{equation}
    \label{eq:single_simple}
    F(t,\lambda)=A\times H(t,\lambda)\Big[1+S(t,\lambda)s\Big]\times C(\lambda)c,
\end{equation}

\noindent where $A$ is the amplitude, $H$ is the baseline SED flux and $s,c$ are scale parameters chosen from a distribution defined for the stretch and color perturbers ($S,C$). 

Equations \ref{eq:salt2}-\ref{eq:byosed} highlight the similarity of \byosed to a PCA-like model such as SALT2, except that \byosed can vary the baseline SED far beyond the flexibility introduced by the $M_1$ and color law components. This more realistic source SED is used in the SNANA simulation as described at the beginning of Section \ref{sec:framework} (and Figure \ref{fig:byosed_schematic}), to produce SN Ia light curves with brightness dependencies beyond the SALT2 framework. By fitting the resulting light curves with SALT2 or other models like those of \citet{saunders_snemo_2018} and \citet{leget_sugar_2020}, we are able to identify model-specific biases and quantify the benefits or drawbacks of a choice of model.

\subsection{Creating \byosed Perturbers from Observables}
\label{sub:perturbers}
The \byosed toolkit accepts perturbers in one of two forms: SED format or a theoretical function. For example, a theoretical function that could be used to create a perturber is from \citet{moreno-raya_dependence_2016}:
\begin{equation}
\label{eq:theoretical_perturber}
\begin{aligned}
M_B(Z)=M_{B,Z_\odot}-... \ \ \ \ \ \ \ \ \ \ \ \ \ \ \ \ \ \ \ \ \ \ \ \ \ \ \ \ \ \ \ \ \ \ \ \ \ \ \ \ \ \ \ \ \\
2.5\log_{10}\Big[1-0.18\frac{Z}{Z_\odot}\Big(1-0.10\frac{Z}{Z_\odot}\Big)\Big]-0.191 \ \rm{mag},
\end{aligned}
\end{equation}
where $Z_\odot$ is the Solar metallicity and $M_B$ is the absolute magnitude of the SN in the B-band. Equation \ref{eq:theoretical_perturber} contains no wavelength dependence, but will simply increase or decrease the brightness of the resulting SN depending on the chosen metallicity of the host galaxy. The equation is based on empirically measured correlations between SN Ia absolute magnitudes and oxygen abundances in their host galaxies by \citet{moreno-raya_dependence_2016}. When paired with a distribution of observed host galaxy metallicities, an SNANA simulation using Equation \ref{eq:theoretical_perturber} and \byosed would produce a SN Ia sample with a theoretically motivated relationship between luminosity and host galaxy metallicity.

\byosed perturbers can also be created from observations. Here we use composite spectra generated from the \kaepora\footnote{\href{https://github.com/msiebert1/kaepora}{https://github.com/msiebert1/kaepora}  \href{https://kaepora.readthedocs.io/en/latest/index.html}{https://kaepora.readthedocs.io}} database, an open-source relational database of SN Ia observations \citep{siebert_investigating_2019}. We use the Gini-weighting method to create the composite spectra, outlined in \citet{siebert_investigating_2019}, which provides a representative spectrum that maximizes S/N ratio while mitigating the impact of high S/N outliers. We control for average properties of phase and $\Delta m_{15} (B)$ to produce sequences of composite spectra with desired properties (e.g. ejecta velocity, host-galaxy mass). We apply selection requirements on $\Delta m_{15} (B)$ to remove peculiar SN Ia spectra from the sample, and to remove correlations between light curve shape and host environment. 

The final SEDs from \kaepora have $2\mbox{\normalfont\AA}$ spectral resolution, and 3-5 day temporal resolution. We perform a linear interpolation in the wavelength dimension as the spacing is more than sufficient to ensure accuracy, but use a more robust Gaussian Process interpolation in the phase dimension to obtain a smoother phase-dependent SED surface. These interpolations are used to sample the composite spectra at the wavelength and phase grid values defined by the baseline SED, and transform them into \byosed perturbers. 

For a summary of how perturbers are created and implemented, see Figure \ref{fig:byosed_schematic} and Section \ref{sub:salt2}. As an example, consider using \kaepora to obtain two composite spectra representing high and low ejecta velocity SN, respectively. \byosed calculates the average of the two composites, $A(t,\lambda)$, and then the fractional difference between the high or low velocity spectra and $A(t,\lambda)$, with respect to the baseline SED:
\begin{equation}
    \label{eq:ex_perturb_H}
    P_{V,H}(t,\lambda)=\frac{V_H(t,\lambda)-A(t,\lambda)}{H(t,\lambda)},
\end{equation}
\begin{equation}
\label{eq:ex_perturb_L}
P_{V,L}(t,\lambda)=\frac{V_L(t,\lambda)-A(t,\lambda)}{H(t,\lambda)},    
\end{equation}
where $P_{V,H}$ ($P_{V,L}$) contains fractional deviations of the composite spectrum $V_H$ ($V_L$) from $A$ for the high (low) velocity composite spectrum, with respect to the baseline SED, $H(t,\lambda)$. The final ``ejecta velocity perturber'' is a function of ejecta velocity as well as phase and wavelength, $P_V(v,t,\lambda)$, defined in the following manner:
\begin{equation}
    \label{eq:ex_perturb_V}
    P_V(v,t,\lambda)=
    \begin{cases} 
      P_{V,H}(t,\lambda) & v>v_0 \\
      P_{V,L}(t,\lambda) & v<v_0, \\
   \end{cases}
\end{equation}
where $v$ is a velocity chosen by \byosed and $v_0$ is the break-point between a high and low ejecta velocity SN. Given a large enough sample of spectra one could produce composites for a range of ejecta velocities instead of just two, and $P_V$ would simply interpolate along the ejecta velocity dimension.  

We use specific perturbers in the course of this paper to show the capabilities of \byosed (see Sections \ref{sec:sims} and \ref{sec:biases}), but of course any choice of perturber is possible and easily included within the \byosed framework. The perturbers in this work were chosen due to their necessity for an accurate SN Ia model (stretch, color), and as a useful starting point to study a subset of the potential biases for cosmology (host mass, SN velocity).

\

\section{Simulating a SN Ia Survey with \byosed and SNANA}
\label{sec:sims}
In this work, we simulate \textit{Roman Space Telescope} SN surveys to demonstrate the value of \byosed, and to test for potential biases in cosmological measurements (Section \ref{sec:biases}). We use the survey and telescope specifications outlined by \citet{hounsell_simulations_2018} (Hereafter \hounsell) to produce realistic light curves and global survey features for the \textit{Roman Space Telescope} mission. For the SED source model we replace SALT2 with a variety of perturbed SEDs generated by \byosed. We first describe the simulation process, then present a series of four simulation case studies.  These examples begin with a Fiducial survey using a baseline SED model with minimal variations, then progressively add more perturbations that modify the underlying SED. All simulations assume a Flat$w$CDM cosmology with $w=-1$ and $\Omega_m=0.315$ \citep{aghanim_planck_2018}.

\subsection{Simulations with SNANA}
\label{sub:snana}
SNANA simulates SN light curves for an arbitrary set of survey properties while accounting for variations in noise, PSF, and cadence. Due to its speed, accuracy, and flexibility, SNANA has become the standard tool for simulating SN surveys in recent years \citep[e.g.,][]{betoule_improved_2014,scolnic_complete_2018,jones_foundation_2019,kessler_first_2019}. Following Figure 1 in \citet{kessler_first_2019}, a brief overview of the SNANA+\byosed simulation is as follows:
\begin{enumerate}
    \item \textbf{Source Model}
    \begin{enumerate}
        \item Generate source SED at each simulated epoch using Baseline SED and \byosed perturber choices (see Sections \ref{sub:sim_stretch_color}-\ref{sub:sim_mass})
        \item Apply cosmological dimming, Galactic extinction, weak lensing, peculiar velocity, and redshift to the SED
        \item Integrate redshifted SED over each filter transmission function to create the noise-free photometric light curve
    \end{enumerate}
    \item \textbf{Noise Model}
    \begin{enumerate}
        \item Use image zero-point to convert each true light curve in magnitude 
    to true flux in 
    photoelectrons.
        \item Compute flux uncertainty from zero-point, PSF and sky noise; apply random fluctuation to true flux.
    \end{enumerate}
    \item \textbf{Trigger Model}
    \begin{enumerate}
        \item Check for detection (SNR $>3\sigma$ in 2 or more bands)
        \item Write selected events to data files
        
    \end{enumerate}
\end{enumerate}

The \textit{Roman} SN survey will provide a unique probe for constraining the nature of dark energy, delivering near-IR observations for thousands of SN Ia over a wide redshift range of $0.01\lesssim z\lesssim3.0$. Since the SALT2 model is poorly constrained at UV ($\lambda<3000$\mbox{\normalfont\AA}) and near-IR ($\lambda>8500$\mbox{\normalfont\AA}) wavelengths, we restrict our simulated redshift range from \textit{Roman} to $0.5 < z < 1.5$, ensuring that our simulations are sampled mostly within the SALT2 wavelength range ($\sim3000-9000$\mbox{\normalfont\AA}). With an updated SN Ia light curve template that extends training into the UV and near-IR, this process could be repeated to evaluate the impact over a wider redshift range.

\hounsell~explored 11 survey strategies including 4 variations on each of 3 primary strategies. Here we investigate only the "All-$z$" primary strategy. The All-$z$ strategy is constructed as a series of ``tiers'' labeled as Shallow, Medium, and Deep. Each tier was simulated here over the redshift range of $0.5-1.5$, using RZYJ filters for the Shallow and Medium tiers, and YJHF for the Deep tier (see \hounsell~Figure 1 for filter transmission details). These tiers have single-visit depths equal to 22.0 mag (J band), 24.8 mag (J band), and 26.2 (H band) for the Shallow, Medium, and Deep tiers respectively.

  \begin{table*}[t]
\centering

\caption{\label{tab:sims} Summary of the simulations created in this section. }
\begin{tabular*}{\textwidth}{@{\extracolsep{\stretch{1}}}*3{l}}
\toprule
  \multicolumn{1}{c}{\textbf{Perturbers}}&\multicolumn{1}{c}{\textbf{Label}}&\multicolumn{1}{c}{\textbf{Section}}\\

\hline
Stretch, color, intrinsic scatter &Fiducial&\ref{sub:sim_stretch_color}\\
Fiducial$+$SED dependence on $SiII$ velocity &Fiducial$+V_{SiII}$\ &\ref{sub:sim_velocity}\\
Fiducial$+$SED dependence on host stellar mass&Fiducial$+M_{\rm{Stellar}}$&\ref{sub:sim_mass}\\
Fiducial$+$SED $z$-dependence on host stellar mass&Fiducial$+M_{\rm{Stellar}}(z)$&\ref{sub:sim_mass}\\

\end{tabular*}

\end{table*}

 The only difference between the H18 All-$z$ simulation and the simulations produced for this work is the choice of rest-frame source SED model as \byosed, and the slightly more limited redshift range. In order to effectively constrain cosmological parameters, we also simulate a low-$z$ anchor sample of SN that mimics the Foundation survey \citep{foley_foundation_2018,jones_foundation_2019}. We proceed with a variety of identical survey simulations, varying only our choice of perturbers in \byosed. In each case, we repeat the full simulation 50 times to create distinct random realizations of the same SN survey, helping us constrain the relevant statistical uncertainties on the biases measured for cosmology. The following sections describe each individual simulation, which are summarized by Table \ref{tab:sims}.

\subsection{Fiducial Simulation}
\label{sub:sim_stretch_color}
In this section, we create our Fiducial simulated SN survey that mimics the H18 All-$z$ simulation, which used SALT2 and its associated stretch and color parameters as the rest-frame source SED model. We add coherent scatter for simplicity across all simulations, which is a good approximation to wavelength-dependent intrinsic scatter models like that of \citet{kessler_testing_2013}. The following sections describe the \byosed perturbers used for the Fiducial simulation, which are meant to produce a SALT2-like source SED (see Equation \ref{eq:single_simple}), and matching of stretch and color to the bias-correction sample (see Section \ref{sub:cosmo_analysis} and Figure \ref{fig:overview}).

\begin{figure*}[t!]
    \centering
    \includegraphics[trim={0cm 13cm 0cm 0cm},clip,width=\textwidth]{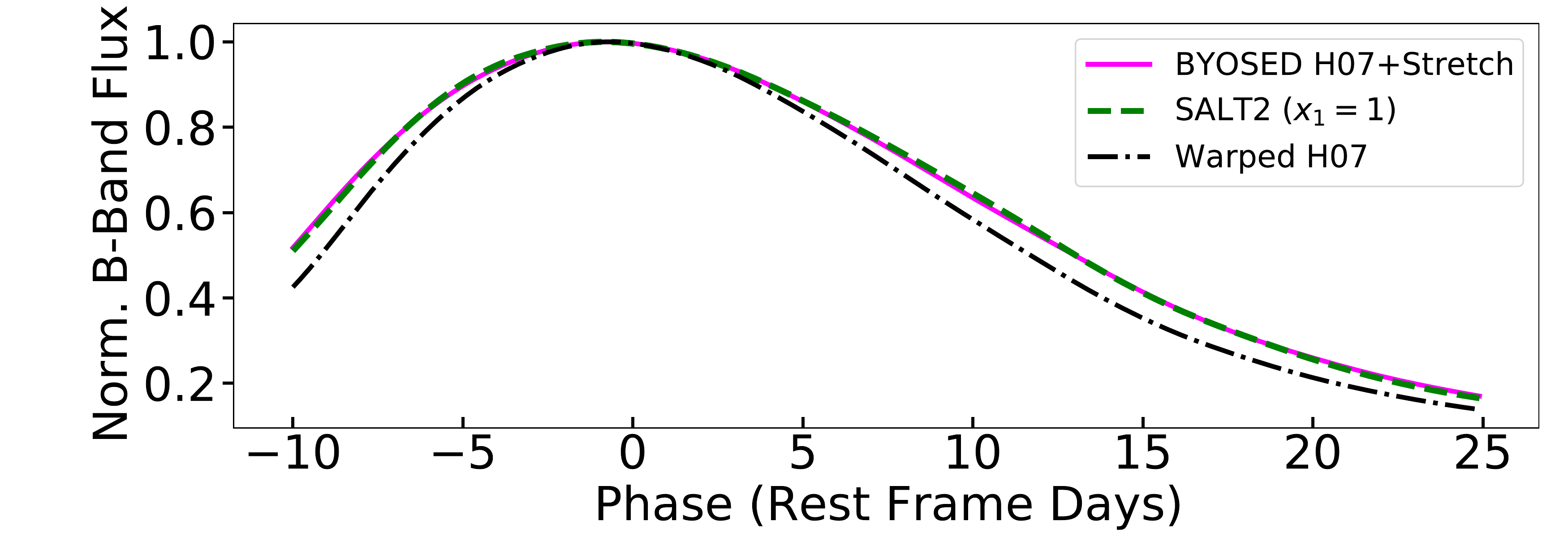}

    \caption{ A light curve generated by the baseline warped H07 model (black dash-dot) compared to a light curve generated by \byosed using the stretch perturber described in this section (magenta solid). A SALT2 light curve for $x1=1$ is shown for comparison (green dashed). }
    \label{fig:stretch}
\end{figure*}

\subsubsection{Stretch Perturber}
\label{sub:stretch}
Any model that attempts to reproduce observed light curves of SN Ia should include some variation of the SED with light curve stretch. For this work, we have implemented stretch as a simple time dilation according to:
\begin{equation}
    \label{eq:stretch}
    F(t,\lambda)=A\times H(t/f,\lambda),
\end{equation}
where $F$ is the final flux, $A$ is a scaling amplitude, $H$ is the baseline SED flux, and $f$ is the stretch factor ($f=1.1$ corresponds to $x_1\sim0$ in the SALT2 framework). Since our analysis includes light curve fitting with the SALT2 model
(Section \ref{sec:biases}), differences between H07 and SALT2 SEDs result in large Hubble residuals that hide the underlying systematic effects from the \byosed perturbers. To focus on the \byosed perturbers, we have
warped the H07 template at each rest-frame epoch so that the synthetic photometry matches that of the SALT2-$M_0$ template. The effect of Equation \ref{eq:stretch} and the H07 warping are shown in Figure \ref{fig:stretch}.

\subsubsection{Color Perturber}
\label{sub:color}
\begin{figure*}[ht!]
    \centering
    \includegraphics[trim={0cm 0cm 0cm 12cm},clip,width=\textwidth]{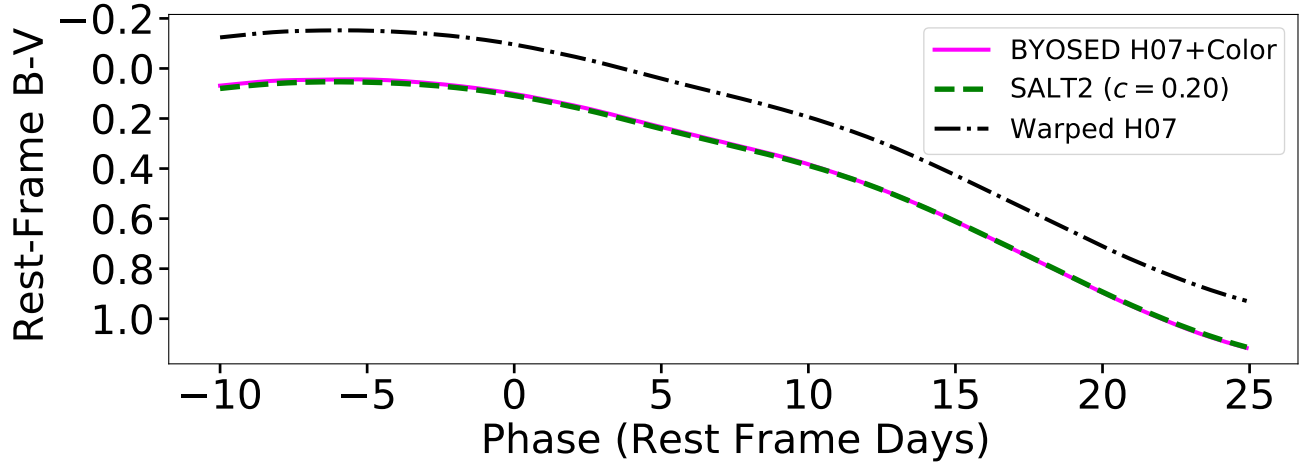}

    \caption{The color curve described by the warped H07 template (black dash-dot) compared with the color curve created by \byosed when the color perturber described in this section is included (magenta solid). The SALT2 color curve with $c=0.2$ is shown for comparison (green dashed).}

    \label{fig:color}
\end{figure*}

As with the stretch perturber in section \ref{sub:stretch}, \byosed needs a color perturber in order to reproduce realistic SN Ia flux variation with wavelength. Again we limit wavelength-dependent differences between models by adopting the SALT2 color law as the \byosed color perturber, implemented as a global correction according to Equation \ref{eq:byosed}. The effect of implementing this \byosed perturber is shown in Figure \ref{fig:color}.

\subsubsection{Matching Stretch and Color Distributions}
\label{sub:salt2_fit}
Since the warped H07 template and stretch perturbers are not the same as the $M_0$ and $M_1$ components of SALT2, we do not expect the simulated values of stretch and color to map precisely onto the measured SALT2 $x_1$ and $c$ parameters, even in the limit of infinite S/N. We generate stretch and color such that the ``measured'' distribution of $x_1,c$ match the measured $x_1,c$ distributions from \citet{scolnic_measuring_2016} (Hereafter SK16) and \hounsell. 
This matching is done iteratively, varying the \byosed stretch and color distributions until the fitted distributions of $c$ and $x_1$ match those of SK16. We determine that distributions are sufficiently ``matched'' when a Kolmogorov-Smirnov (KS) test identifies the distributions as identical, with a confidence of $>95\%$.

\subsubsection{Matching Luminosity Parameters}
\label{sub:lum_match}
In addition to the matching of stretch and color distributions in Section \ref{sub:salt2_fit}, we undergo a similar process for choosing the stretch and color luminosity coefficients for \byosed. These coefficients ($\alpha\prime$ and $\beta\prime$) are analogs of $\alpha$ and $\beta$ in the SALT2 model, and are used in the \byosed simulation to adjust the SN magnitude by $\alpha\prime b_s-\beta\prime b_c$, with $b_s,b_c$ the \byosed stretch and color parameters. Therefore, we vary $\alpha\prime$ and $\beta\prime$ in \byosed until $\alpha$ and $\beta$ measured in the analysis (Section \ref{sub:cosmo_analysis}) are within $3\sigma$ of the values used by \hounsell~ ($\alpha=0.14,\beta=3.1$). For the Fiducial simulation, we find that input values of $\alpha\prime=0.068, \ \beta\prime=3.08$ yield average measurements of $\alpha=0.142\pm0.001, \ \beta=3.10\pm0.01$ across the 50 survey iterations. We note that $\beta\prime\sim\beta$ as expected because we are using the SALT2 color law as the \byosed perturber, but $\alpha\prime$ and $\alpha$ differ by a factor of $\sim2$ because of our choice of stretch perturber. While we could avoid the process outlined here and in section \ref{sub:salt2_fit} by simply using the SALT2 $M_0$ and $M_1$ templates as our baseline SED and stretch perturber, these transformations provide some valuable independence from SALT2 in our simulations. Past analyses assumed that the model used for light curve fitting perfectly explains the variation introduced by simulations, and we wish to avoid that here as much as possible.

\subsection{Fiducial$+V_{SiII}$ Simulation }
\label{sub:sim_velocity}
We next follow the process outlined in Section \ref{sub:sim_stretch_color}, and include a SN velocity perturber in addition to the color and stretch perturbers (Sections \ref{sub:stretch}-\ref{sub:color}). Following the methodology outlined in Section \ref{sub:lum_match}, we find that \byosed input values of $\alpha\prime=0.07, \ \beta\prime=3.32$ yield average BBC measurements of $\alpha=0.142\pm0.001, \ \beta=3.11\pm0.01$. Here, we simulate each SN with a dependency (SN velocity) that has no corresponding model component in SALT2. Ejecta velocities are simulated according to the observed population from \citet{siebert_possible_2020}.

\begin{figure*}[t!]
    \centering

    \includegraphics[trim={2cm 0cm 3cm 0cm}, clip,width=\textwidth]{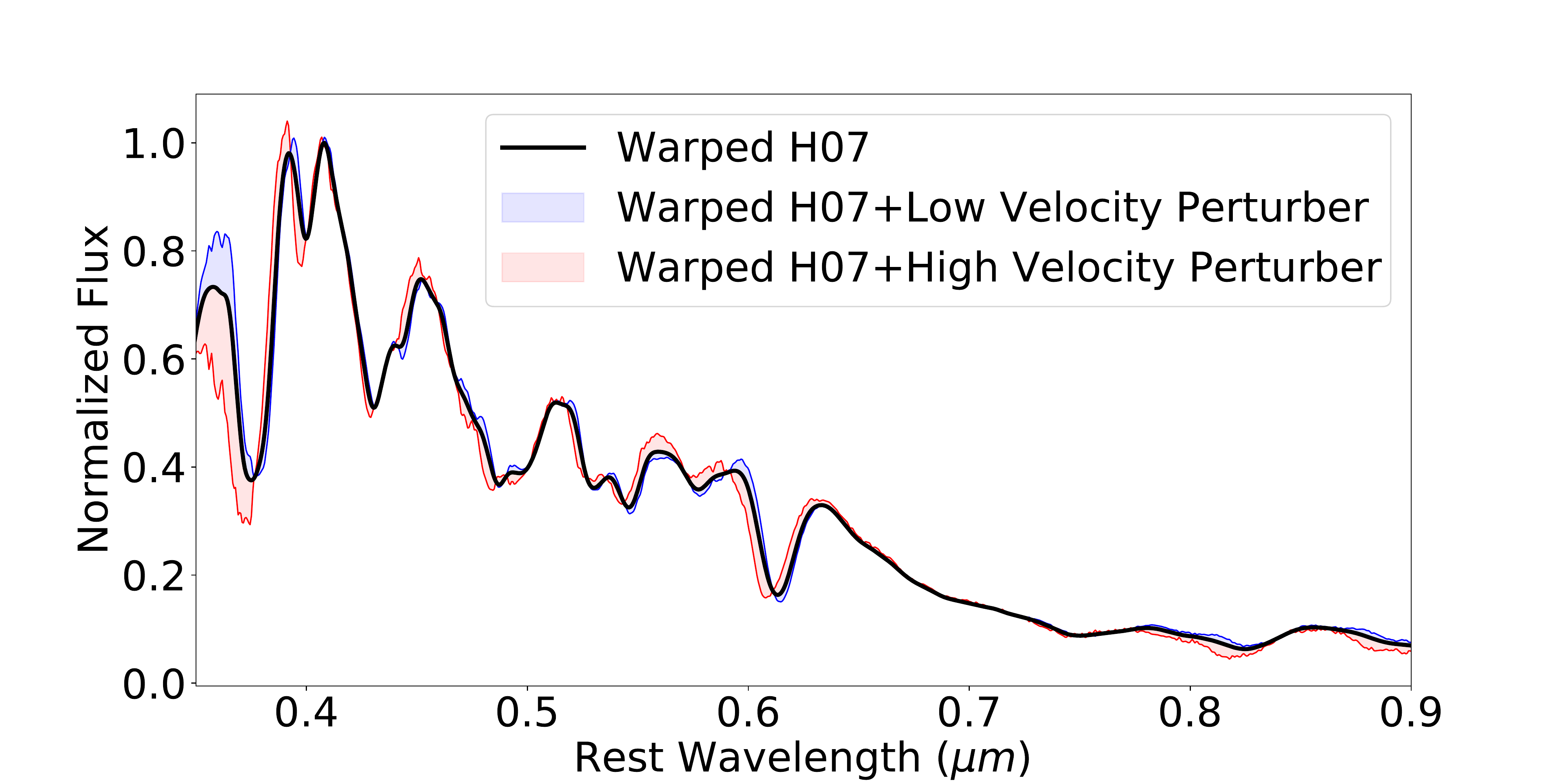}

    \caption{The flux at peak for the SEDs produced by the addition of a low- (blue) or high-velocity (red) perturber in the \byosed framework. The perturbers invoke variations in flux as a function of phase and wavelength relative to the baseline SED, here the warped H07 template (black). The scale of each velocity perturber has been increased by a factor of 2 to more clearly illustrate their relative differences. }
    
    \label{fig:vel_spec}
\end{figure*}

\subsubsection{Ejecta Velocity Perturber}
\label{sub:velocity}
\begin{figure}[h!]
    \centering

    \includegraphics[trim={1cm 0cm 0cm 0cm},clip,width=.5\textwidth]{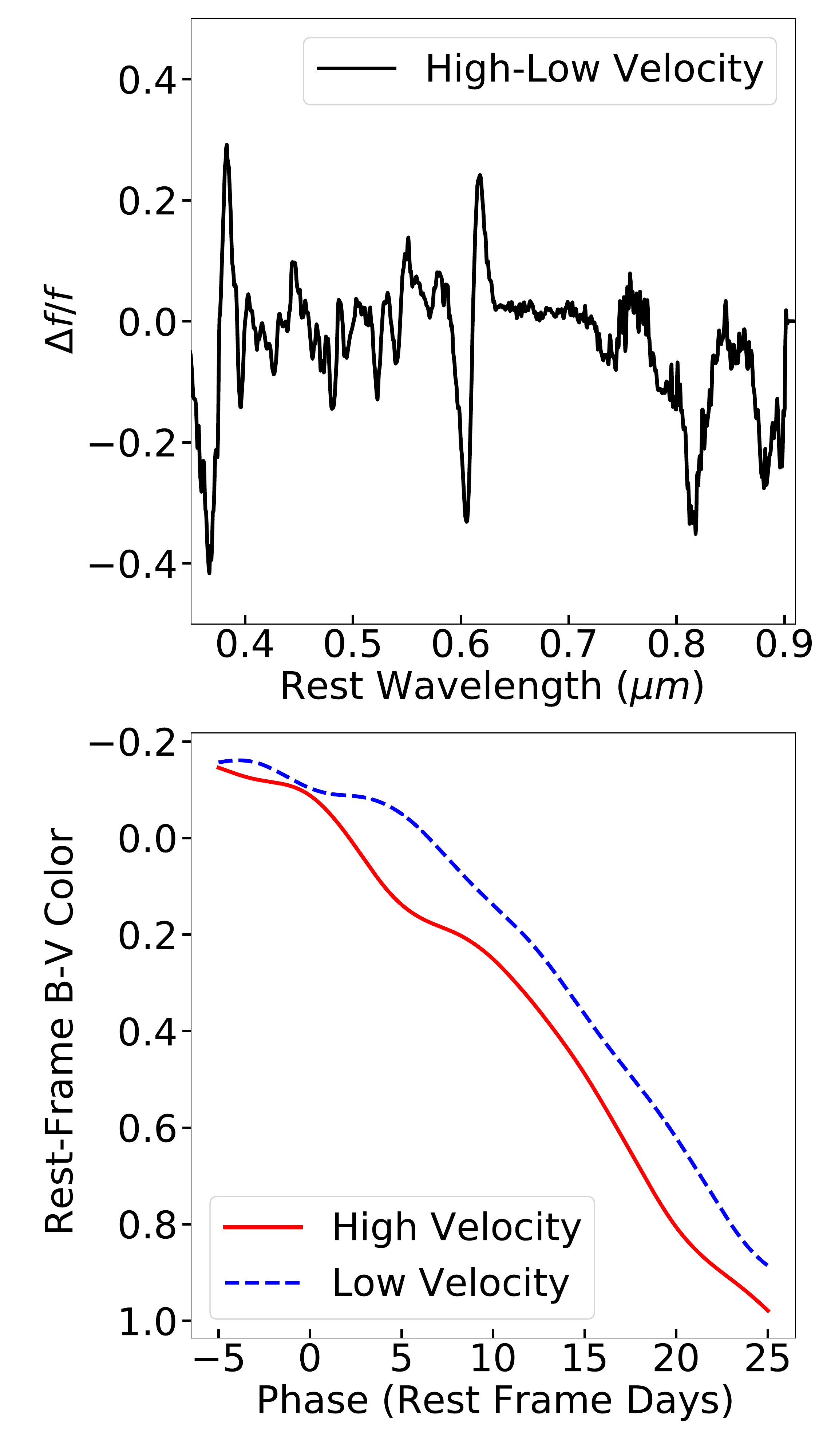}

    \caption{(Top) The fractional difference between the \kaepora composite spectrum for high and low ejecta velocity at peak brightness, relative to the warped H07 template. (Bottom) The rest-frame B-V color curve for low (blue dashed) and high (red solid) ejecta velocity. }
    
    \label{fig:velocity}
\end{figure}

SN velocity is an active topic of investigation in the SN Ia community, with hints that high and low-velocity SN Ia vary in their intrinsic colors and intrinsic color scatter \citep[e.g.,][]{foley_velocity_2011,foley_relation_2012,mandel_type_2014,pan_type_2015,siebert_possible_2020}. This correlation motivated the inclusion of a velocity perturber in our simulations. We use the \kaepora database to generate an SED time series (spectral sequence, see Section \ref{sub:perturbers}) for low- and high-velocity SNe Ia defined by $v_{\rm{Si}{II}} > -11{,}000$ and $v_{\rm{Si}{II}} < -11{,}000$, respectively. Each of these sequences comprises 20 composite spectra generated with phase bin sizes of 3 days ranging from $-14$ to $+47$ days and resulting in a median $\Delta m_{15} (B)$ of $1.10\pm0.01$ and $1.13\pm0.02$ for low- and high-velocity, respectively. By translating these composite spectra into a perturber, \byosed is able to produce simulated SN Ia with spectra and light curves matching those of low- or high-ejecta velocity SN (Figure \ref{fig:vel_spec}). Figure \ref{fig:velocity} shows low- and high-velocity SN Ia color curves using \byosed. \byosed-simulated SN Ia with lower velocity are bluer than those with high velocity, consistent with previous observations \citep{foley_velocity_2011,siebert_possible_2020}. We treat these color differences as ``intrinsic'' (i.e., not correlated with luminosity), as proposed by \citet{foley_velocity_2011}.

\subsection{Fiducial$+M_{Stellar}$ and Fiducial$+M_{Stellar}(z)$ Simulations}
\label{sub:sim_mass}
As a final example, we include host galaxy mass correlations as a \byosed perturber in our simulations, in addition to the stretch and color perturbers outlined in Section \ref{sub:sim_stretch_color}. Host galaxy masses are simulated according to the observed population from \citet{siebert_investigating_2019}. In this section we create two sets of simulations: First we include host mass as a static \byosed perturber with no redshift evolution. In the analysis, a standard static ``host mass step'' will correct for the perturber. A small $w$-bias is expected because the host mass perturber is based upon spectra from the \kaepora database, and thus we attempt to correct a wavelength-dependent perturber with a single wavelength-independent correction to the Hubble residuals. Following Section \ref{sub:lum_match}, we find that input \byosed values of $\alpha\prime=0.07, \ \beta\prime=3.14$ yield BBC measurements of $\alpha=0.143\pm0.001, \ \beta=3.12\pm0.01$.

In the second simulation we introduce a redshift-dependence using a linear approximation of the mass-step vs. redshift relationship proposed by \citet{childress_ages_2014}. In the analysis, we use the same host mass step correction that is applied to the first simulation. Here input \byosed parameters of $\alpha\prime=0.068, \ \beta\prime=3.14$ yield BBC measurements of $\alpha=0.142\pm0.001, \ \beta=3.12\pm0.01$.

\subsubsection{Host Galaxy Mass Perturber}
\label{sub:hostmass}

Previous work has found that high-mass galaxies hosting SN Ia tend to have negative Hubble residuals, motivating the simulations produced in this section \citep[e.g.,][]{kelly_hubble_2010,sullivan_dependence_2010,lampeitl_effect_2010}. We again use \kaepora to generate low- and high-mass host galaxy spectral sequences using spectra from SNe Ia with $\log M_{\rm{Stellar}}<10.7$ and $\log M_{\rm{Stellar}}>10.7$, respectively. This choice of host mass step maximizes the coverage of \kaepora spectra in each mass range. Each of these sequences comprises 15 composite spectra ranging from $-10$ to $+62$ days. In order to include a statistically significant sample in each bin, phase bin sizes range from 4 days at early times to 7 days at later times. Given the correlation between SN~Ia luminosity and host-galaxy environment \citep{hamuy_search_2000, howell_progenitors_2001}, we select $\Delta m_{15} (B)$ bins such that each composite spectrum has a median $\Delta m_{15} (B)$ of $1.15\pm0.06$ mag and $1.16\pm0.06$ mag for low- and high-mass, respectively. Thus, the remaining differences in the composite spectra (if any) should be related to host galaxy mass and not light curve shape. 

\begin{figure}[ht!]
    \centering
    \includegraphics[trim={1cm 0cm 0cm 0cm},clip,width=.5\textwidth]{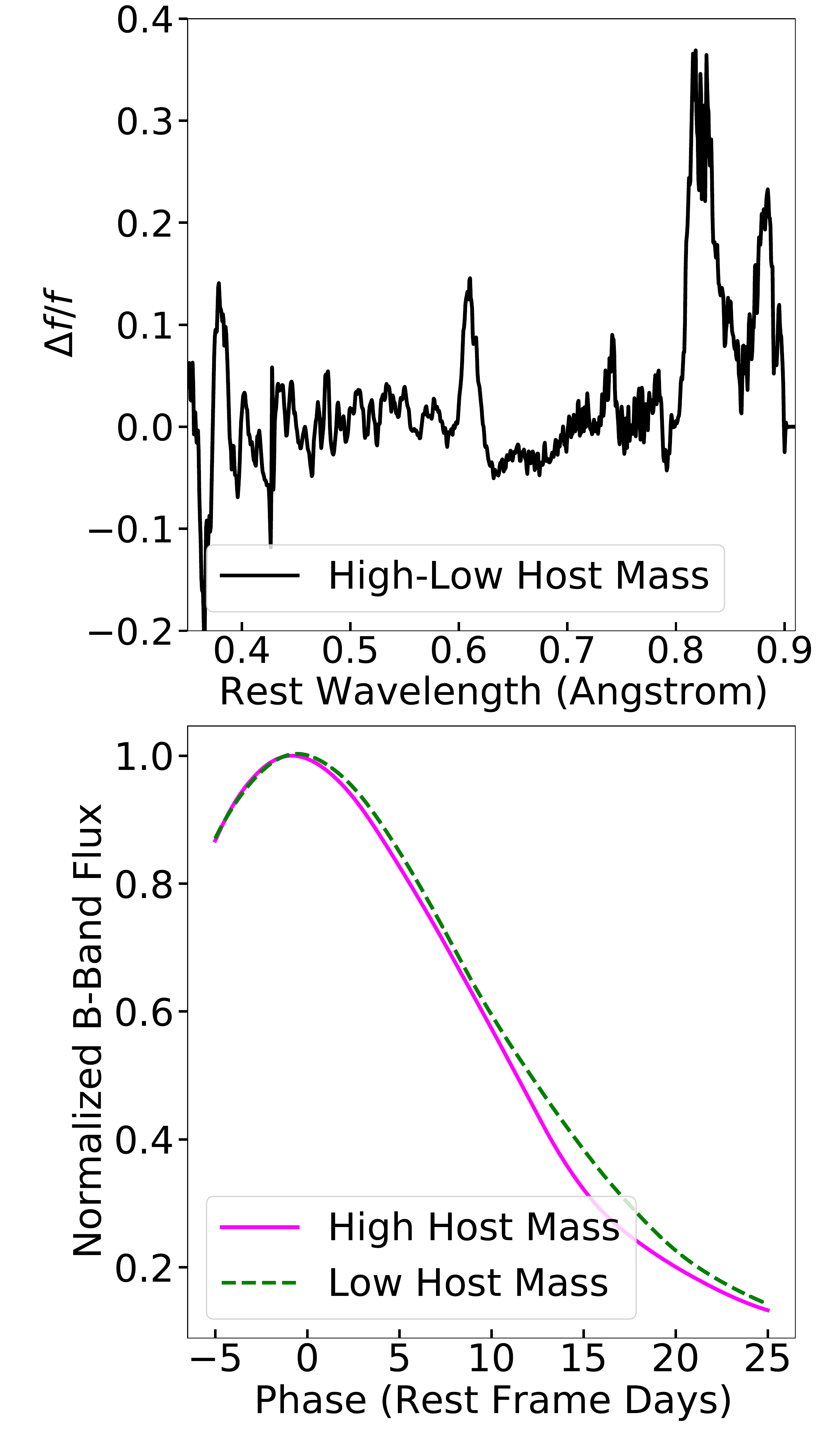}
    
    \caption{(Top) The fractional difference between the \kaepora composite spectrum for high and low host galaxy mass as peak brightness, relative to the warped H07 template. (Bottom) Broad-band light curves based on the SED described by \byosed using the warped H07 template with low (green dashed) and high (magenta solid) host galaxy mass perturbers.}
    
    \label{fig:mass}
\end{figure}

We translate the composite spectra into a \byosed host mass perturber in the manner described in Section \ref{sub:perturbers}. Figure \ref{fig:mass} shows a SN Ia light curve and spectrum from a low- and high- mass host galaxy using \byosed. The similar B-band light curves reflect the selections based on $\Delta m_{15}$ above, and the resulting mass step should be due to variations in SED features alone \citep{siebert_investigating_2019}. We do observe some residual differences in recovered $x_1$ even after controlling for $\Delta m_{15} (B)$ that are not present in the case of the velocity composites (Section \ref{sub:velocity}). It is possible that controlling for $x_1$ instead of $\Delta m_{15} (B)$ would reduce these differences, but we have chosen not to control for SALT2 parameters at the \kaepora stage so that the perturbers remain as general as possible. We cannot be sure that our perturber exactly mirrors the effect of the mass step in nature, but regardless our implementation of the $z-$dependence of this step faithfully follows the theoretical prediction from \citet{childress_ages_2014} and the low-significance measurement from \citet{scolnic_complete_2018}. 

\begin{figure*}[t!]
    \centering
    \includegraphics[width=\textwidth]{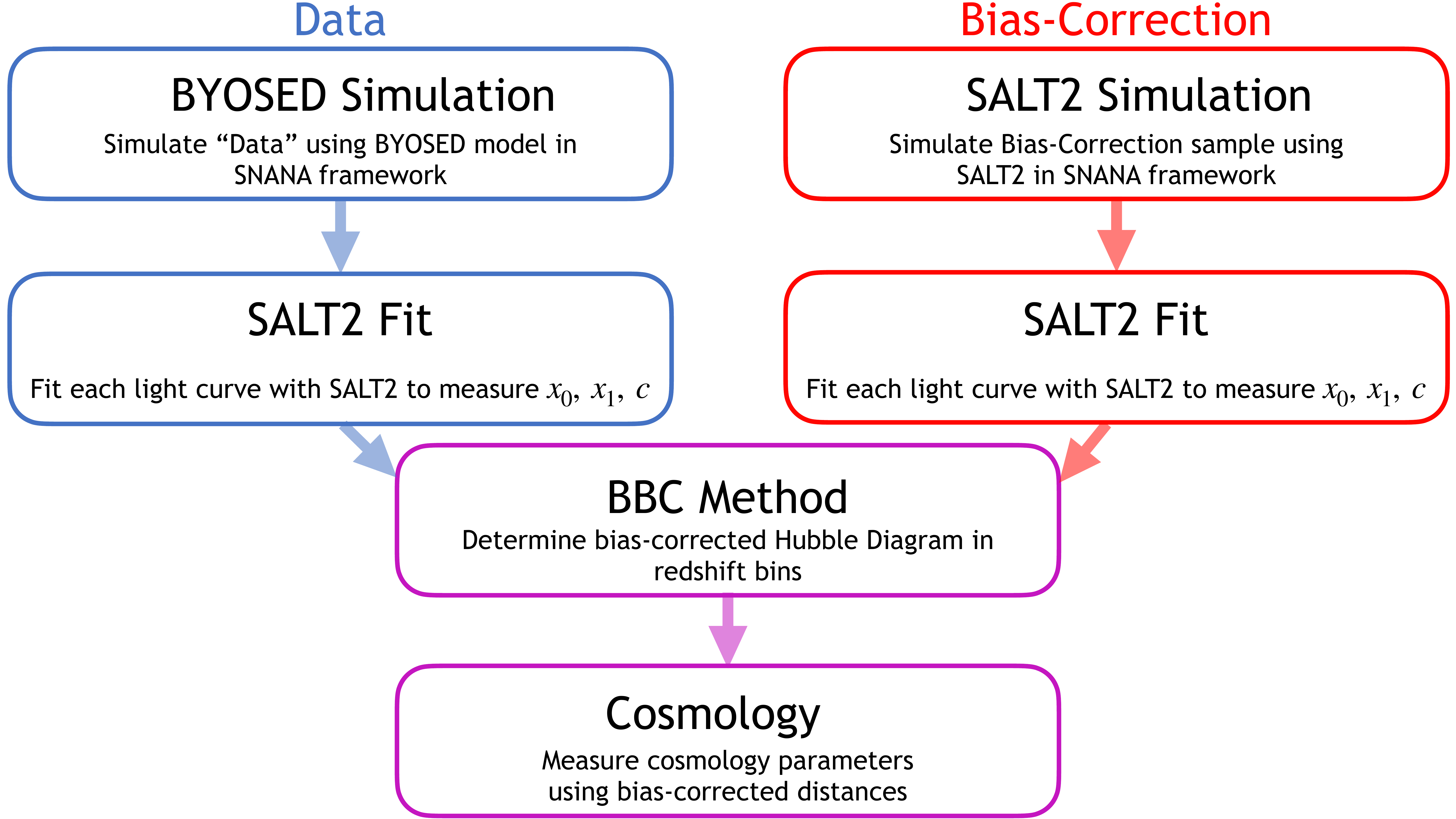}
    \caption{Overview of the analysis process.}
    \label{fig:overview}
\end{figure*}

\section{Identifying Potential Sources of Bias for \textit{Roman}}
\label{sec:biases}
For each choice of \byosed perturber(s), we have created 50 simulated \textit{Roman Space Telescope} SN Ia surveys, as well as the anchor datasets of low-$z$ SN Ia. All simulated light curves are fit with the SALT2 model. The fitted SALT2 parameters for the simulations in Section \ref{sec:sims} and the bias-correction sample (see Section \ref{sub:cosmo_analysis}) are used by the BBC method to determine bias-corrected distance moduli in 30 redshift bins. In this work we define measured cosmological biases with respect to variations in the ``$w$'' parameter in the Flat$w$CDM model, whereby the dark energy equation of state is defined as $p=w\rho c^2$. This process is outlined in Figure \ref{fig:overview}, and details of the final two stages are given in Section \ref{sub:cosmo_analysis}. The cosmological measurements are presented in Section \ref{sub:cosmo_results}.

\subsection{Analysis}
\label{sub:cosmo_analysis}
We transform fitted SALT2 light curve parameters into distances by way of a modified Tripp formula \citep{tripp_two-parameter_1998}:
\begin{equation}
\label{eq:tripp}
\mu=m_B-M+\alpha x_1-\beta c+\gamma G_{\rm{Host}}+\Delta\mu_{\rm{Bias}},
\end{equation}
where $\mu$ is the distance modulus, $\alpha$ ($\beta$) is the coefficient of relation between SN Ia luminosity and stretch (color), and $M$ is the peak absolute magnitude of an $x_1=c=0$ SN Ia assuming some nominal value of $H_0$. $\gamma$ corrects for a dependence on host galaxy stellar mass, with $G_{\rm{Host}}=1/2$ or $-1/2$ if $M_{\rm{Stellar}}>10^{10.7}M_\odot$ or $M_{\rm{Stellar}}<10^{10.7}M_\odot$, respectively. Finally, the $\Delta\mu_{\rm{Bias}}$ term is a selection bias correction determined by BBC, described below. Recall that in this work, the mass step location is set at $10^{10.7}M_\odot$ instead of the typical value of $10^{10}M_\odot$ to maximize the number of \kaepora spectra corresponding to high- and low- mass host galaxies (Section \ref{sub:hostmass}). For the Fiducial and Fiducial$+V_{SiII}$ simulations, where we do not associate a simulated host galaxy mass to each SN, we simply set the value of $\gamma$ to $0$ at the BBC stage.

We then employ the BEAMS with Bias Corrections (BBC) method \citep{kunz_bayesian_2007,kunz_beams_2013,kessler_correcting_2017}, which estimates the correction term in Equation \ref{eq:tripp}, $\Delta\mu_{\rm{Bias}}$, based on a large simulated sample of SN Ia in a 5D space of \{$z,x_1,c,\alpha,\beta$\}. BBC produces a bias-corrected Hubble Diagram, from which we measure $w$, and requires two primary inputs:
\begin{enumerate}
    \item \textbf{Data}: The ``observed'' light curves, here simulated by SNANA using \byosed as the source SED (see Sections \ref{sec:framework}-\ref{sec:sims}). These light curves are fit with the SALT2 model.
    \item \textbf{Bias-correction sample}: A large sample of SN Ia light curves simulated and fit with the SALT2 model. Note that here the ``data'' and bias-correction sample are both fit with the SALT2 model, but simulated with independent models.
\end{enumerate}
One must ensure that the properties of the SN in the ``data'' are well-matched by the large simulations, which determine the nuisance parameters $\alpha$ and $\beta$ and simultaneously identify bias corrections for the remaining parameters in the Tripp formula: $x_1,c$. The iterative process outlined in Section \ref{sub:salt2_fit} ensures that the fitted light curve parameter distributions for the simulated SN (i.e. $x_1,c$) match the distributions of our large simulations for BBC, defined in SK16. Here we are proceeding in the same manner as previous cosmological analyses \citep[e.g.,][]{scolnic_complete_2018}, except that here the ``data'' are generated by SNANA with \byosed as the source model. The large bias-correction simulations ($\sim10$x larger than those with \byosed) are generated with a nominal SALT2 simulation in line with \citet{kessler_correcting_2017}. While we use the SALT2 model for our bias-correction simulations to mimic previous work and identify biases, we note that replacing the SALT2 model with \byosed removes the biases reported in Section \ref{sub:cosmo_results}, as expected.

For the four choices of \byosed perturbers, we implement the BBC method for each of the 50 distinct simulations created in Section \ref{sec:sims} (a total of 200 survey simulations). This method creates a set of BBC distance modulus measurements, $\mu_{BBC,z}$, which are bias-corrected. We use the BBC redshift, distance pairs in 30 redshift bins to fit a cosmological model using flat priors of $\Omega_M=0.315\pm0.007$ \citep{aghanim_planck_2018} and $w=-1.0\pm1.0$, minimizing the difference $\mu_{BBC,z}-\mu_{model,z}$. 

\begin{table*}[t!]
\centering
\caption{\label{tab:w_shifts} Cosmological bias test results. }
\begin{tabular*}{\textwidth}{@{\extracolsep{\stretch{1}}}*6{r}}
\toprule
 \multicolumn{1}{c}{Simulation} & \multicolumn{1}{c}{$w^a$} 
    & \multicolumn{1}{c}{$\langle\sigma_w\rangle^b$}& 
    \multicolumn{1}{c}{$\Delta w_{\rm{Bias}}^c$}&
    \multicolumn{1}{c}{$|\Delta w_{\rm{Bias}}|/\langle\sigma_w\rangle_{Fid}$}&
    \multicolumn{1}{c}{$\gamma^d$}\\
\hline
Fiducial &$-0.986\pm 0.004^e$&$0.020$&0&0&$-$ \\
Fiducial$+V_{SiII}$ &$-1.009\pm0.006$&$0.022$&$-0.023\pm0.006$&1.11&$-$\ \\
Fiducial$+M_{\rm{Stellar}}$&$-0.980\pm0.005$&$0.029$&$0.006\pm0.007$&0.28&$0.071\pm0.003$ \\
Fiducial$+M_{\rm{Stellar}}(z)$ &$-0.965\pm0.005$&$0.021$&$0.021\pm0.006$&1.02&$0.043\pm0.003$ \\
\hline
\end{tabular*}
\footnotesize
\begin{flushleft}
$^a$ Uncertainties are the standard error on the mean (SEM).

$^b$ The median of the statistical uncertainties measured for the 50 distinct survey iterations.

$^c$ Uncertainties are the RMSE between the Fiducial survey and each respective survey.

$^d$ Weighted average of BBC-measured $\gamma$ for all 50 survey iterations.

$^e$ Recall that this measurement of $w$ is not expected or required to be $-1$ (the input cosmology), and biases in column 4 are measured relative to this value.
\end{flushleft}
\end{table*}

\normalsize

\subsection{Results}
\label{sub:cosmo_results}
\begin{figure*}[ht!]
    \vspace{-10pt}
    \includegraphics[trim={0cm 1.5cm 3cm 3cm}, clip,width=.9\textwidth]{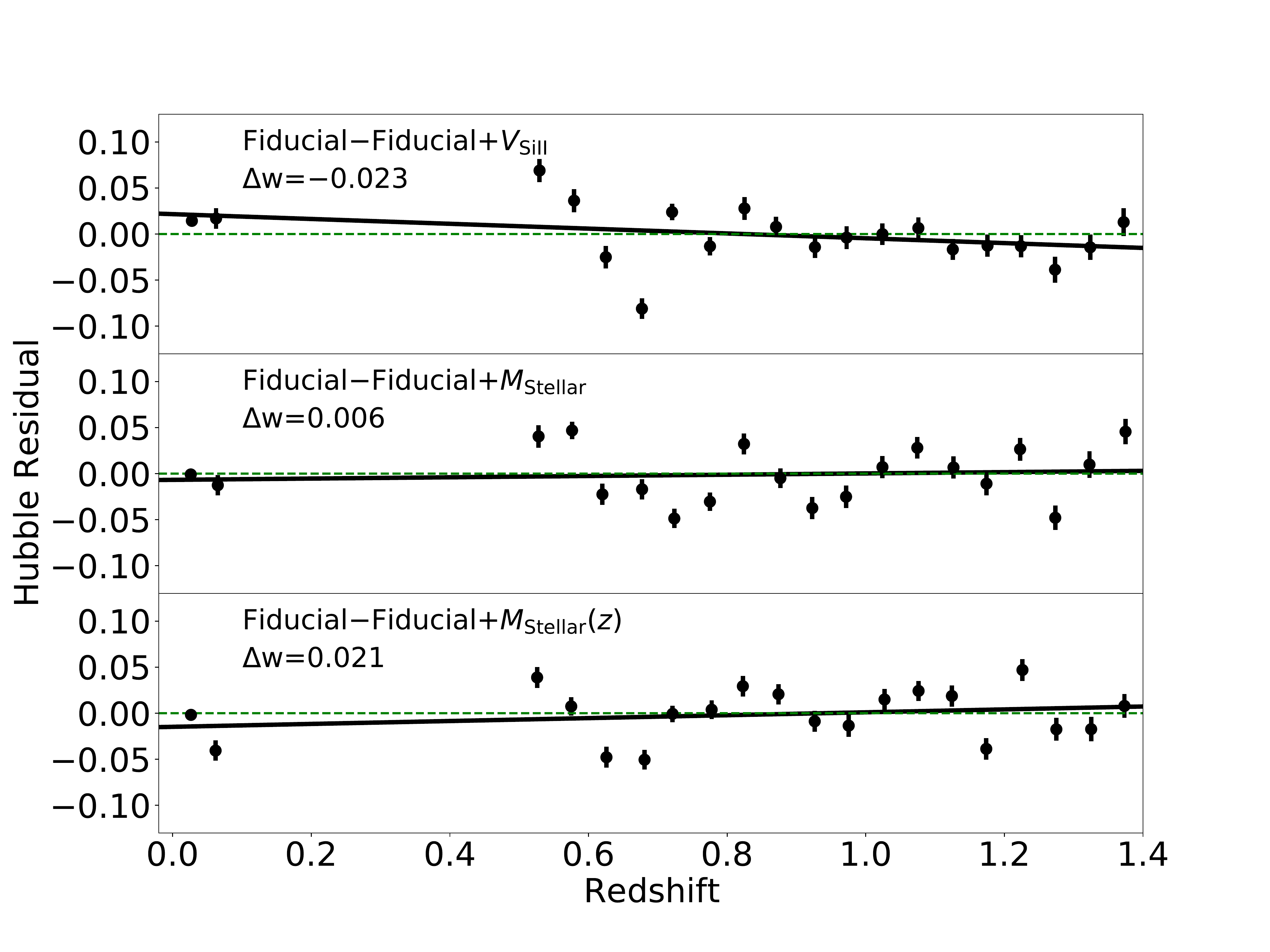}
    \centering
    \caption{Example of the difference in Hubble residuals between the Fiducial survey and each respective survey investigated in this work.}
    
    \label{fig:hr_stack}
\end{figure*}

The analysis of each simulation gives us a unique constraint on $w$, which are combined with a weighted average to obtain final measurements for the Fiducial, Fiducial$+V_{SiII}$, Fiducial$+M_{\rm{Stellar}}$, and Fiducial$+M_{\rm{Stellar}}(z)$ simulations. Biases in $w$ relative to the Fiducial simulation ($\Delta w_{\rm{Bias}}=w-w_{\rm{Fiducial}}$) are reported in Table \ref{tab:w_shifts}. We note that the Fiducial simulation does not precisely recover the input cosmology of $w=-1$ due to the wavelength-dependent differences between our baseline warped H07 template combined with a stretch perturber and the SALT2 model. These differences introduce small variations in distance measurement accuracy as a function of redshift that reveal themselves as a biased measurement on $w$. By comparing the subsequent measurements of $w$ to this Fiducial survey, we are directly quantifying the impact of the additional perturbers. Figure \ref{fig:hr_stack} shows the difference in Hubble residuals between the Fiducial survey and each respective survey.

\begin{figure*}[ht!]
    \vspace{-10pt}
    \includegraphics[width=.9\textwidth]{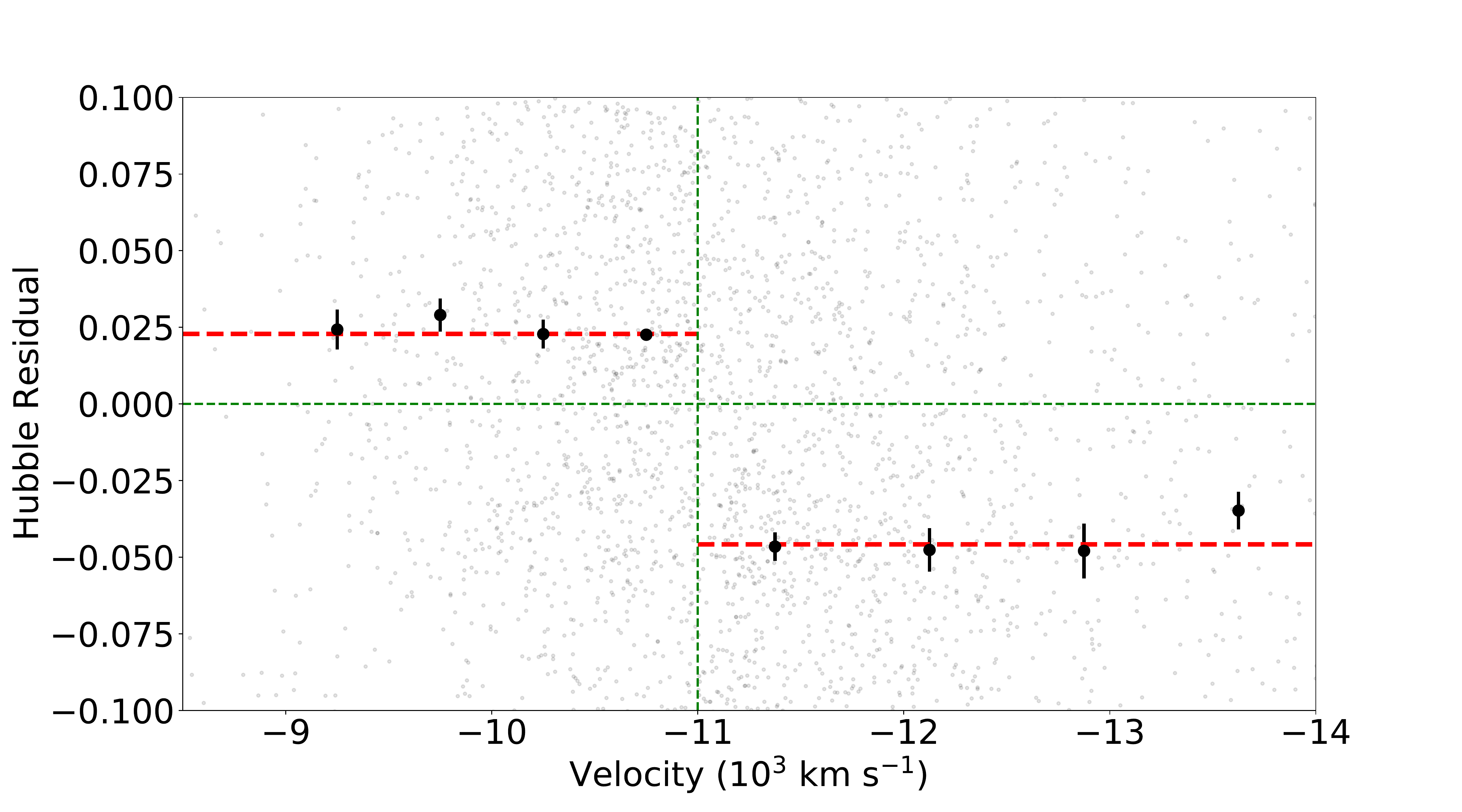}
    \centering
    \caption{Example of binned Hubble residuals as a function of ejecta velocity. The input \byosed simulation used $-11\times10^{3}$ km s$^{-1}$ as the cutoff between ``high'' and ``low'' velocity SN, and the resulting ``step'' is $\sim0.07$ mag, consistent with previous work \citep[][$0.091\pm0.035$ mag]{siebert_possible_2020}. Error bars show the standard error on the mean, and dashed red lines show the separate means for high and low mass hosts.}
    
    \label{fig:step_vel}
\end{figure*}

\begin{figure*}[ht!]
    \vspace{-10pt}
    \includegraphics[width=.9\textwidth]{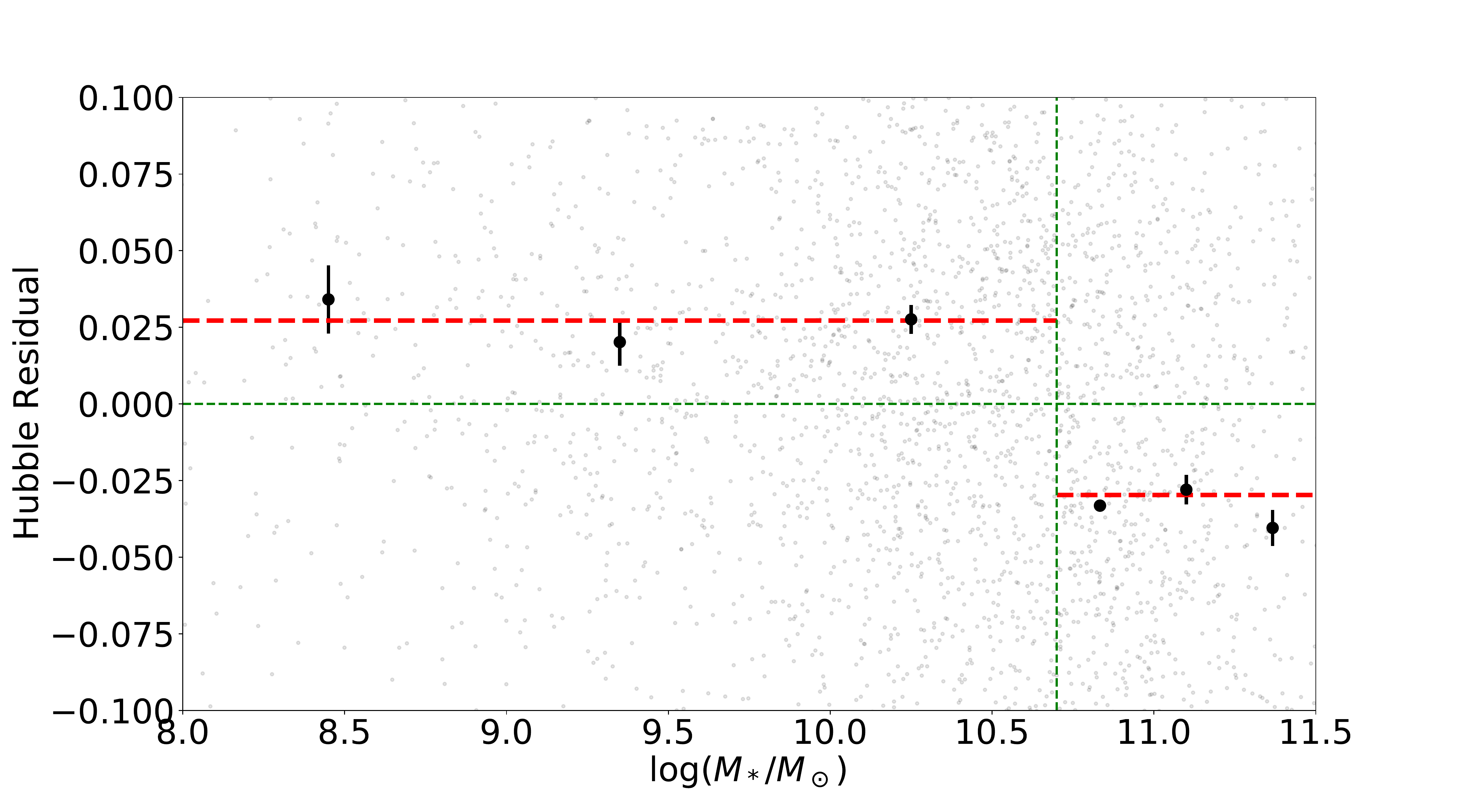}
    \centering
    \caption{Example of binned Hubble residuals as a function of global host galaxy mass. The input \byosed simulation used 10.7 as the cutoff between ``high'' and ``low'' mass hosts, and the resulting ``step'' is $\sim0.071$ mag, consistent with previous work. Error bars show the standard error on the mean, and dashed red lines show the separate means for high and low mass hosts.}
    
    \label{fig:step_mass}
\end{figure*}

\begin{figure*}[ht!]
    \includegraphics[width=.9\textwidth]{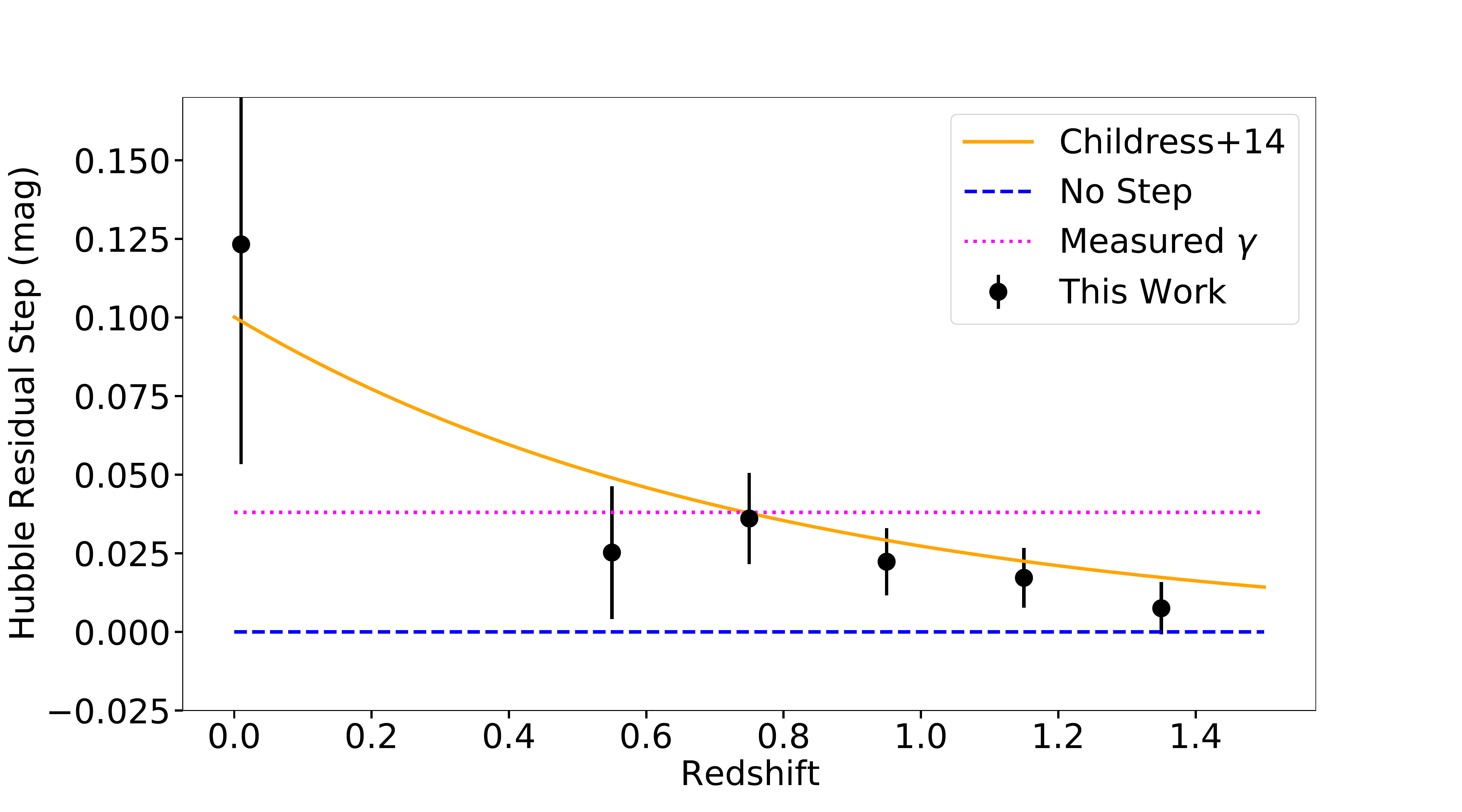}
    \centering
    \caption{Example of the evolution of the host mass step with redshift. The solid orange line represents a prediction of the evolution from \citet{childress_ages_2014}. The ``observed'' step as a function of redshift from this work are shown as black points with error bars, while the measured host mass step (independent of redshift) is shown as a dotted magenta line. A scenario with no host mass step would be represented by the dashed blue line.}
    \label{fig:mass_z}
\end{figure*}

For the Fiducial$+V_{SiII}$ case, we find a bias on $w$ of $-0.023\pm0.006$ with respect to the Fiducial survey cosmology, which is $1.11$ times the statistical uncertainty on $w$ for the Fiducial survey. We also observe a Hubble residual step of $\sim0.07$ mag consistent with \citet{siebert_possible_2020} ($0.091\pm0.035$ mag), which is unsurprising as both studies used the same \kaepora spectra (Figure \ref{fig:step_vel}). For the host mass simulations, we have broken the results into the simulations that do not include redshift evolution, and those that include redshift evolution. For these cases, as we are simulating meaningful host galaxy masses associated with each SN, we measure the $\gamma$ parameter for each simulation. We find that without redshift evolution (Fiducial$+M_{\rm{Stellar}}$) a host mass step is observed (Figure \ref{fig:step_mass}), and a simple luminosity correction with $\gamma=0.071\pm0.003$ results in a very small bias on $w$ ($0.006\pm0.007$) that is  $0.28$ times the Fiducial statistical uncertainty.  When redshift evolution is included for the host mass perturber (Fiducial$+M_{\rm{Stellar}}(z)$), we find a bias in $w$ of $0.021\pm0.006$, or $\sim1.02$ times the Fiducial statistical uncertainty. The measured host mass step correction term ($\gamma=0.043\pm0.003$) under-corrects for the effect at low-redshift, and over-corrects at high redshift due to the choice of redshift evolution (Figure \ref{fig:mass_z}).

\pagebreak

\pagebreak

\section{Discussion and Conclusions}
\label{sec:discussion}
In this work, we have presented a new open-source tool, \byosed, that will enable simple and rigorous tests of potential biases in SN Ia luminosity distance measurements. \byosed is already included in the commonly used SNANA software package, and can be employed immediately to investigate properties of the SN Ia cosmological measurement system that could impact cosmological analyses. \byosed is capable of producing simulations using perturbers based on theoretical models or observed spectra. 

As a proof of concept, we have applied \byosed to simulate a representative \textit{Roman Space Telescope} SN Ia survey. We created a fiducial population of SN Ia analogous to simulations in the literature that use the SALT2 model exclusively, and then compared the dark energy equation of state parameter ($w$)  inferred by the Fiducial study, to three further cases. The results are summarized as follows:

\

\

\

\

\

\

\

\

\

\

\

\begin{enumerate}
    \item \textbf{Fiducial$\mathbf{+V_{SiII}}$}
    \begin{enumerate}
        \item The addition of a SN velocity perturber causes a bias in $w$ relative to the Fiducial case, likely due to the wavelength dependence of the velocity perturber causing redshift-dependent variation in measured color.
        \item The measured bias on $w$, relative to the Fiducial simulation, is $-0.023\pm0.006$. This is 1.11 times the statistical uncertainty on $w$ for the Fiducial survey, indicating that this could be a concerning source of potential systematic uncertainty for the \textit{Roman} survey.
    \end{enumerate}
    \item \textbf{Fiducial$\mathbf{+M_{Stellar}}$}
    \begin{enumerate}
        \item SN properties are strongly correlated with host galaxy mass; while we have attempted to control for shape differences between SNe in high- and low-mass hosts, there could be residual correlations causing the step to be driven by shape and color differences in the perturbers that are treated as intrinsic.  Therefore it is unclear if the bias in this case is modeled in a way representative of the true host mass step, but serves as a useful test of a possible effect of host mass-based SED variations.
        \item The BBC method is able to correct for this effect, and we are left with a small bias on $w$ of 0.006$\pm0.007$ relative to the Fiducial survey. This is 0.28 times the statistical uncertainty on $w$ for the Fiducial survey, indicating this would be a sub-dominant source of systematic uncertainty.
    \end{enumerate}
    \textbf{Fiducial$\mathbf{+M_{Stellar}(z)}$}
    \begin{enumerate}
        \item The addition of redshift-evolution causes the simple host mass step correction to be insufficient, with SN Ia under-corrected at low-$z$ but over-corrected at high-$z$
        \item We observe a bias of 0.021$\pm0.006$ on $w$ relative to the Fiducial simulation, or $\sim1.02$ times the Fiducial statistical uncertainty. This suggests a redshift-evolving host mass correlation could contribute a systematic uncertainty on par with the statistical uncertainty on $w$ if a linear trend were not constrained by the cosmological analysis.
    \end{enumerate}
\end{enumerate}

Based on these results, there could be an impact from these possible investigated relationships on the \textit{Roman} SN Ia survey. These new potential systematics are roughly equal to the projected statistical uncertainty on measurements of $w$ with the \textit{Roman Space Telescope} \citep{hounsell_simulations_2018}, making them important considerations when outlining uncertainty budgets for cosmology. Future work should investigate potential changes to the \textit{Roman} SN Ia survey strategy that could mitigate, or help us understand, such effects. For example, it is plausible that a survey optimized for low-$z$ discoveries will be impacted quite differently than one optimized for high-$z$ discoveries when considering perturbers that vary with redshift. Alternatively a well-balanced survey covering a wide redshift range or that includes high-S/N spectroscopy of many SN Ia might be the most effective at identifying and understanding distance biases, as well as the extent to which SN Ia distance measurement systematics vary with redshift. Spectroscopic measurements could also help delineate between the effects that truly impact cosmological measurements, and those that do not manifest themselves in observed data.

 \byosed can be used to investigate potentially redshift-dependent effects such as host galaxy mass, metallicity, and star formation rate that could impact the results from different survey strategies in unpredictable ways.  This work represents the first tool capable of flexibly producing light curve simulations containing observed correlations between SED features and the SN progenitor or host environment. In much the same way that large-scale simulations with a model like SALT2 have improved our knowledge of SN Ia systematics over the last decade, \byosed can provide the next step forward in probing systematic uncertainties that we have thus far been unable to accurately simulate at the SED-level. \byosed can be integrated into existing cosmology pipelines to more accurately represent these systematics for cosmological measurements, and test the relative accuracy of different SN Ia light curve models.

The \byosed tool will enable many interesting investigations into any perturber that has been proposed to impact SN Ia distance measurements. Even from this preliminary study into the effect of SN velocity and host galaxy mass, we have gleaned useful information about the importance of acknowledging and accounting for these factors. More realistic simulations with \byosed will also lead to more accurately trained light curve models by identifying systematics, and help leverage the next generation of SN Ia surveys as one of our most important cosmological probes. 

\

\noindent \acknowledgments

\noindent This  work  was  supported by 
the National Aeronautics and Space Administration (NASA)
Headquarters under the NASA Future Investigators in Earth and Space Science and Technology (FINESST) award 80NSSC19K1414, and by NASA  contract  No.\  NNG17PX03C  issued  through  the \textit{Roman} Science Investigation Teams Program. Support for this work was also provided by NASA through grant No.\ HST-AR-15808 from the Space Telescope Science Institute, which is operated by AURA, Inc., under NASA contract NAS 5-26555. The UCSC team is supported in part by NASA grant NNG17PX03C, NSF grant AST-1815935, the Gordon \& Betty Moore Foundation, the Heising-Simons Foundation, and by a fellowship from the David and Lucile Packard Foundation to R.J.F. M.R.S. is supported by the National Science Foundation Graduate Research Fellowship Program under grant No.\ 1842400. DS is supported by DOE grant DE-SC0010007 and the David and Lucile Packard Foundation.  D.O.J. is supported by a Gordon and Betty Moore Foundation postdoctoral fellowship at the University of California, Santa Cruz and by NASA through the NASA Hubble Fellowship grant HF2-51462.001 awarded by the Space Telescope Science Institute, which is operated by the Association of Universities for Research in Astronomy, Inc., for NASA, under contract NAS5-26555. MD is supported by the Horizon Fellowship at the Johns Hopkins University. This work at Rutgers University was supported by NASA contract NNG16PJ34C issued through the Roman (WFIRST) SIT program.

\pagebreak

\bibliographystyle{aas}

\end{document}